\documentclass{article}

\begin{document}

\title{Majorana Neutrino: Chirality and Helicity}    
\author{Valeriy V. Dvoeglazov\\
Universidad de Zacatecas, M\'exico}               
\date{August 8, 2011}          
\maketitle

\abstract{We introduce the Majorana spinors in the momentum representation. 
They obey the Dirac-like equation with eight components, which has been first introduced by Markov.
Thus, the Fock space for corresponding quantum fields is doubled (as shown by Ziino).
Particular attention has been paid to the questions of chirality and helicity 
(two concepts which frequently are confused in the literature) for 
Dirac and Majorana states.}

\newpage

{\large

\section{The Dirac Equation.}


The Dirac equation has been considered in detail in a pedagogical way~\cite{Sakurai,Ryder}:
\begin{equation}
[i\gamma^\mu \partial_\mu -m]\Psi (x) =0\,.\label{Dirac}
\end{equation}
At least, 3 methods of its derivation exist:
\begin{itemize}
  \item the Dirac one (the Hamiltonian should be linear in $\partial/\partial x^i$, and be compatible with $E^2 -{\bf p}^2 c^2 =m^2 c^4$);
  \item the Sakurai one (based on the equation \linebreak $(E- {\bf \sigma} \cdot {\bf p}) (E+ {\bf \sigma} \cdot {\bf p}) \phi =m^2 \phi$);
  \item the Ryder one (the relation between  2-spinors at rest is $\phi_R ({\bf 0}) = \pm \phi_L ({\bf 0})$).
\end{itemize}
The $\gamma^\mu$ are the Clifford algebra matrices 
\begin{equation}
\gamma^\mu \gamma^\nu +\gamma^\nu \gamma^\mu = 2g^{\mu\nu}\,.
\end{equation}
Usually, everybody uses the following definition of the field operator~\cite{Itzykson}:
\begin{equation}
\Psi (x) = \frac{1}{(2\pi)^3}\sum_\sigma \int \frac{d^3 {\bf p}}{2E_p} [ u_\sigma ({\bf p}) a_\sigma ({\bf p}) e^{-ip\cdot x}
+ v_\sigma ({\bf p}) b_\sigma^\dagger ({\bf p})] e^{+ip\cdot x}]\,,
\end{equation}
as given {\it ab initio}, while the general scheme has been given by~\cite{Bogoliubov}.

I studied in the previous 
works~\cite{Dvoeglazov1,Dvoeglazov2,Dvoeglazov3}:
\begin{itemize}
  \item $\sigma \rightarrow h$  (the helicity basis); the motivation was that Berestetskii, Lifshitz and Pitaevskij~\cite{BLP}
claimed that the helicity eigenspinors are {\it not} the parity eigenspinors, and {\it vice versa};
  \item  the modified Sakurai derivation (the additional $m_2 \gamma^5$ term may appear in the Dirac equation);
  \item  the derivation of the Barut equation~\cite{Barut} from the first principles, namely 
  based on the generalized Ryder relation, ($\phi_L^h ({\bf 0}) = \hat A \phi_L^{-h\,\ast} ({\bf 0}) + \hat B \phi_L^{h\,\ast} ({\bf 0})$). In fact, we have the second mass state ($\mu$-meson)  from that equation:
  \begin{equation}[i\gamma^\mu \partial_\mu - \alpha \partial_\mu \partial^\mu  /m -\beta] \psi =0\,\end{equation}  
due to the existence of {\it two} parameters;
\item the self/anti-self charge-conjugate Majorana 4-spinors\linebreak 
\cite{Majorana, Bilenky} in the momentum representation.
  \end{itemize}

The Wigner rules~\cite{Wigner} of the Lorentz transformations
for the $(0,S)$ left- $\phi_L ({\bf p})$ and  the $(S,0)$ right-
$\phi_R ({\bf p})$ spinors are:
\begin{eqnarray}
(S,0):\phi_R ({\bf p})= \Lambda_R ({\bf p} \leftarrow
{\bf 0})\,\phi_R ({\bf 0})  =  \exp (+\,{\bf S} \cdot
{\bf \varphi}) \,\phi_R ({\bf 0}),\\
(0,S):\phi_L ({\bf p}) = \Lambda_L ({\bf p} \leftarrow
{\bf 0})\,\phi_L
({\bf 0})  =  \exp (-\,{\bf S} \cdot {\bf \varphi})\,\phi_L
({\bf 0}),\label{boost0}
\end{eqnarray}
with ${\bf \varphi} = {\bf n} \varphi$ being the boost parameters:
\begin{eqnarray}
&&cosh (\varphi ) =\gamma = \frac{1}{\sqrt{1-v^2/c^2}}, \\
&&sinh (\varphi ) =\beta \gamma =\frac{v/c}{\sqrt{1-v^2/c^2}},\, 
tanh (\varphi ) =v/c\,.
\end{eqnarray} 
They are well known and given, {\it e.g.}, in~\cite{Wigner,Faustov,Ryder}. 

On using the Wigner boost rules and the Ryder relations we can recover the Dirac equation in the matrix form:
\begin{eqnarray}
\pmatrix{\mp m \, 1 & p_0 + {\bf \sigma}\cdot {\bf p}\cr
p_0 - {\bf \sigma}\cdot {\bf p} & \mp m \, 1\cr} \psi (p^\mu)\,
=\, 0\,,
\end{eqnarray}
or
$(\gamma\cdot p - m) u ({\bf p})=0$ and $(\gamma\cdot p + m) v ({\bf p})=0$.
We have used the property $\left [\Lambda_{L,R} ({\bf p}
\leftarrow {\bf 0})\right ]^{-1} =
\left [\Lambda_{R,L} ({\bf p} \leftarrow {\bf 0})\right ]^\dagger$ above,
and that both ${\bf S}$ and $\Lambda_{R,L}$ are Hermitian
for the finite $(S=1/2,0)\oplus (0,S=1/2)$ representation
of the Lorentz group.
Introducing $\psi (x) \equiv \psi (p)  \exp (\mp ip\cdot x)$
and letting $p_\mu \rightarrow i\partial_\mu$, the above equation
becomes the Dirac equation (\ref{Dirac}).

The solutions of the Dirac equation are denoted by \linebreak $u ({\bf p}) = column (\phi_R ({\bf p})\quad \phi_L ({\bf p}))$ and $v ({\bf p}) =\gamma^5 u ({\bf p})$.  Let me remind
that the boosted 4-spinors in the common-used basis  (the standard representation of $\gamma$ matrices) are
\begin{eqnarray}
u_{{1\over 2},{1\over 2}} &=& \sqrt{\frac{(E+m)}{2m}}
\pmatrix{1\cr 0\cr p_z/(E+m)\cr p_r/(E+m)\cr}\,,\nonumber\\
u_{{1\over 2},-{1\over 2}} &=&\sqrt{\frac{(E+m)}{2m}}
\pmatrix{0\cr 1\cr p_l/(E+m)\cr -p_z/(E+m)\cr}\,,\\
v_{{1\over 2},{1\over 2}} &=& \sqrt{\frac{(E+m)}{2m}}\pmatrix{p_z/(E+m)\cr p_r/(E+m)\cr
1\cr 0\cr}\,,\nonumber\\
v_{{1\over 2},-{1\over 2}} &=&\sqrt{\frac{(E+m)}{2m}} \pmatrix{p_l/(E+m)\cr -p_z/(E+m)\cr 0\cr 
1\cr}\,.
\end{eqnarray}
$E=\sqrt{{\bf p}^2 +m^2}>0$, $p_0=\pm E$, $p^\pm = E\pm p_z$, $p_{r,l}= p_x\pm ip_y$.
They are the eigenstates of the helicity in the case $p_3 =\vert {\bf p}\vert$ only.
They  are the parity eigenstates with the eigenvalues of $\pm 1$. In
the parity operator the matrix $\gamma_0^{chiral}=\pmatrix{1&0\cr 0 &
-1\cr}$ was used as usual. They also describe
eigenstates of the charge operator, $Q$, if at rest
\begin{equation}\label{rb}
\phi_R ({\bf 0})
=\pm \phi_L ({\bf 0})
\end{equation}
(otherwise the corresponding physical states are no longer
the charge eigenstates). 
Their normalizations are:
\begin{eqnarray}
&&\bar u_\sigma ({\bf p})u_{\sigma^\prime} ({\bf p}) =+\delta_{\sigma\sigma^\prime}\,,\\
&&\bar v_\sigma ({\bf p})v_{\sigma^\prime} ({\bf p}) =-\delta_{\sigma\sigma^\prime}\,,\\
&&\bar u_\sigma ({\bf p})v_{\sigma^\prime} ({\bf p}) = 0\,.
\end{eqnarray}
The bar over the 4-spinors signifies the Dirac conjugation.

Thus, in the most papers one uses the basis for charged particles in the $(S,0)\oplus (0,S)$ representation (in general) 
\begin{eqnarray}
&&\hspace{-25mm}u_{+\sigma} ({\bf 0}) = N_\sigma\pmatrix{1\cr 0\cr . \cr .
\cr . \cr 0\cr},\,
u_{\sigma-1} ({\bf 0})=N_\sigma \pmatrix{0\cr
1\cr . \cr .  \cr . \cr 0\cr},\ldots
v_{-\sigma} ({\bf 0})=N_\sigma \pmatrix{0\cr 0\cr . \cr . \cr . \cr 1\cr}\nonumber\\
&&
\end{eqnarray}
Sometimes, the normalization factor is convenient to choose $N(\sigma)=m^\sigma$ in order
the rest spinors to vanish in the massless limit. 

However, other constructs are possible in the $(1/2,0)\oplus (0,1/2)$ representation~\cite{Barut,SenGupta, Tokuoka,Raspini,Ziino,Ahluwalia,Dv-ff}.

\section{Majorana Spinors in the Momentum Representation.}

During the 20th century various authors introduced {\it self/anti-self} charge-conjugate 4-spinors
(including in the momentum representation), see~\cite{Majorana,Bilenky,Ziino,Ahluwalia}. 
Later \cite{Lounesto,Dvoeglazov1,Dvoeglazov2,Kirchbach,Rocha} {\it etc} studied these spinors, they found corresponding dynamical equations, gauge transformations 
and other specific features of them.
The definitions are:
\begin{equation}
C= e^{i\theta} \pmatrix{0&0&0&-i\cr
0&0&i&0\cr
0&i&0&0\cr
-i&0&0&0\cr} {\cal K} = -e^{i\theta} \gamma^2 {\cal K}
\end{equation}
is the anti-linear operator of the charge conjugation. ${\cal K}$ is the complex conjugation operator. We  define the {\it self/anti-self} charge-conjugate 4-spinors 
in the momentum space
\begin{eqnarray}
C\lambda^{S,A} ({\bf p}) &=& \pm \lambda^{S,A} ({\bf p})\,,\\
C\rho^{S,A} ({\bf p}) &=& \pm \rho^{S,A} ({\bf p})\,,
\end{eqnarray}
where
\begin{equation}
\lambda^{S,A} (p^\mu)=\pmatrix{\pm i\Theta \phi^\ast_L ({\bf p})\cr
\phi_L ({\bf p})}\,,\label{lambda}
\end{equation}
and
\begin{equation}
\rho^{S,A} ({\bf p})=\pmatrix{\phi_R ({\bf p})\cr \mp i\Theta \phi^\ast_R ({\bf p})}\,.
\label{rho}
\end{equation}
The Wigner matrix is
\begin{equation}
\Theta_{[1/2]}=-i\sigma_2=\pmatrix{0&-1\cr
1&0}\,,
\end{equation}
and $\phi_L$, $\phi_R$ can be boosted with $\Lambda_{L,R}$ 
matrices.\footnote{Such definitions of 4-spinors differ, of course, from the original Majorana definition in x-representation:
\begin{equation}
\nu (x) = \frac{1}{\sqrt{2}} (\Psi_D (x) + \Psi_D^c (x))\,,
\end{equation}
$C \nu (x) = \nu (x)$ that represents the positive real $C-$ parity field operator only. However, the momentum-space Majorana-like spinors 
open various possibilities for description of neutral  particles 
(with experimental consequences, see~\cite{Kirchbach}). For instance,``for imaginary $C$ parities, the neutrino mass 
can drop out from the single $\beta $ decay trace and 
reappear in $0\nu \beta\beta $, a curious and in principle  
experimentally testable signature for a  non-trivial impact of 
Majorana framework in experiments with polarized sources."}

The rest $\lambda$ and $\rho$ spinors can be defined conforming with (\ref{lambda},\ref{rho}) 
in analogious way with the Dirac spinors:
\begin{eqnarray}
\lambda^S_\uparrow ({\bf 0}) &=& \sqrt{\frac{m}{2}}
\pmatrix{0\cr i \cr 1\cr 0}\,,\,
\lambda^S_\downarrow ({\bf 0})= \sqrt{\frac{m}{2}}
\pmatrix{-i \cr 0\cr 0\cr 1}\,,\,\\
\lambda^A_\uparrow ({\bf 0}) &=& \sqrt{\frac{m}{2}}
\pmatrix{0\cr -i\cr 1\cr 0}\,,\,
\lambda^A_\downarrow ({\bf 0}) = \sqrt{\frac{m}{2}}
\pmatrix{i\cr 0\cr 0\cr 1}\,,\,\\
\rho^S_\uparrow ({\bf 0}) &=& \sqrt{\frac{m}{2}}
\pmatrix{1\cr 0\cr 0\cr -i}\,,\,
\rho^S_\downarrow ({\bf 0}) = \sqrt{\frac{m}{2}}
\pmatrix{0\cr 1\cr i\cr 0}\,,\,\\
\rho^A_\uparrow ({\bf 0}) &=& \sqrt{\frac{m}{2}}
\pmatrix{1\cr 0\cr 0\cr i}\,,\,
\rho^A_\downarrow ({\bf 0}) = \sqrt{\frac{m}{2}}
\pmatrix{0\cr 1\cr -i\cr 0}\,.
\end{eqnarray}
Thus, in this basis, with the appropriate normalization (``mass dimension"),
the explicite forms of the 4-spinors of the second kind  $\lambda^{S,A}_{\uparrow\downarrow}
({\bf p})$ and $\rho^{S,A}_{\uparrow\downarrow} ({\bf p})$
are
\begin{eqnarray}
&&\hspace{-15mm}\lambda^S_\uparrow ({\bf p}) = \frac{1}{2\sqrt{E+m}}
\pmatrix{ip_l\cr i (p^- +m)\cr p^- +m\cr -p_r},
\lambda^S_\downarrow ({\bf p})= \frac{1}{2\sqrt{E+m}}
\pmatrix{-i (p^+ +m)\cr -ip_r\cr -p_l\cr (p^+ +m)}\nonumber\\
\\
&&\hspace{-15mm}\lambda^A_\uparrow ({\bf p}) = \frac{1}{2\sqrt{E+m}}
\pmatrix{-ip_l\cr -i(p^- +m)\cr (p^- +m)\cr -p_r},
\lambda^A_\downarrow ({\bf p}) = \frac{1}{2\sqrt{E+m}}
\pmatrix{i(p^+ +m)\cr ip_r\cr -p_l\cr (p^+ +m)}\nonumber\\
\\
&&\hspace{-15mm}\rho^S_\uparrow ({\bf p}) = \frac{1}{2\sqrt{E+m}}
\pmatrix{p^+ +m\cr p_r\cr ip_l\cr -i(p^+ +m)},
\rho^S_\downarrow ({\bf p}) = \frac{1}{2\sqrt{E+m}}
\pmatrix{p_l\cr (p^- +m)\cr i(p^- +m)\cr -ip_r}\nonumber\\
\\
&&\hspace{-15mm}\rho^A_\uparrow ({\bf p}) = \frac{1}{2\sqrt{E+m}}
\pmatrix{p^+ +m\cr p_r\cr -ip_l\cr i (p^+ +m)},
\rho^A_\downarrow ({\bf p}) = \frac{1}{2\sqrt{E+m}}
\pmatrix{p_l\cr (p^- +m)\cr -i(p^- +m)\cr ip_r}.\nonumber
\\
\end{eqnarray}
As claimed by~\cite{Ahluwalia} $\lambda$ and $\rho$ 4-spinors are {\it not} the eigenspinors of the helicity.\footnote{See next Sections for the discussion.} Moreover, 
$\lambda$ and $\rho$ are NOT the eigenspinors of the parity, as opposed to the Dirac case (in this representation $P=\pmatrix{0&1\cr 1&0}R$, where $R= ({\bf x} \rightarrow -{\bf x})$).
The indices $\uparrow\downarrow$ should be referred to the chiral helicity 
quantum number introduced 
in the 60s, $\eta=-\gamma^5 h$, Ref.~\cite{SenGupta}.
While 
\begin{equation}
Pu_\sigma ({\bf p}) = + u_\sigma ({\bf p})\,,
Pv_\sigma ({\bf p}) = - v_\sigma ({\bf p})\,,
\end{equation}
we have
\begin{equation}
P\lambda^{S,A} ({\bf p}) = \rho^{A,S} ({\bf p})\,,
P \rho^{S,A} ({\bf p}) = \lambda^{A,S} ({\bf p})\,,
\end{equation}
for the Majorana-like momentum-space 4-spinors on the first quantization level.
In this basis one has also the relations between the above-defined 4-spinors:
\begin{eqnarray}
\rho^S_\uparrow ({\bf p}) \,&=&\, - i \lambda^A_\downarrow ({\bf p})\,,\,
\rho^S_\downarrow ({\bf p}) \,=\, + i \lambda^A_\uparrow ({\bf p})\,,\,\\
\rho^A_\uparrow ({\bf p}) \,&=&\, + i \lambda^S_\downarrow ({\bf p})\,,\,
\rho^A_\downarrow ({\bf p}) \,=\, - i \lambda^S_\uparrow ({\bf p})\,.
\end{eqnarray}

The normalizations of the spinors $\lambda^{S,A}_{\uparrow\downarrow}
({\bf p})$ and $\rho^{S,A}_{\uparrow\downarrow} ({\bf p})$ are the following ones:
\begin{eqnarray}
\overline \lambda^S_\uparrow ({\bf p}) \lambda^S_\downarrow ({\bf p}) \,&=&\,
- i m \quad,\quad
\overline \lambda^S_\downarrow ({\bf p}) \lambda^S_\uparrow ({\bf p}) \,= \,
+ i m \quad,\quad\\
\overline \lambda^A_\uparrow ({\bf p}) \lambda^A_\downarrow ({\bf p}) \,&=&\,
+ i m \quad,\quad
\overline \lambda^A_\downarrow ({\bf p}) \lambda^A_\uparrow ({\bf p}) \,=\,
- i m \quad,\quad\\
\overline \rho^S_\uparrow ({\bf p}) \rho^S_\downarrow ({\bf p}) \, &=&  \,
+ i m\quad,\quad
\overline \rho^S_\downarrow ({\bf p}) \rho^S_\uparrow ({\bf p})  \, =  \,
- i m\quad,\quad\\
\overline \rho^A_\uparrow ({\bf p}) \rho^A_\downarrow ({\bf p})  \,&=&\,
- i m\quad,\quad
\overline \rho^A_\downarrow ({\bf p}) \rho^A_\uparrow ({\bf p}) \,=\,
+ i m\quad.
\end{eqnarray}
All other conditions are equal to zero.

The dynamical coordinate-space equations are:\footnote{Of course, the signs at the mass terms
depend on, how do we associate the positive- or negative- frequency solutions with $\lambda$ and $\rho$.}
\begin{eqnarray}
i \gamma^\mu \partial_\mu \lambda^S (x) - m \rho^A (x) &=& 0 \,,
\label{11}\\
i \gamma^\mu \partial_\mu \rho^A (x) - m \lambda^S (x) &=& 0 \,,
\label{12}\\
i \gamma^\mu \partial_\mu \lambda^A (x) + m \rho^S (x) &=& 0\,,
\label{13}\\
i \gamma^\mu \partial_\mu \rho^S (x) + m \lambda^A (x) &=& 0\,.
\label{14}
\end{eqnarray}
Neither of them can be regarded as the Dirac equation.
However, they can be written in the 8-component form as follows:
\begin{eqnarray}
\left [i \Gamma^\mu \partial_\mu - m\right ] \Psi_{_{(+)}} (x) &=& 0\,,
\label{psi1}\\
\left [i \Gamma^\mu \partial_\mu + m\right ] \Psi_{_{(-)}} (x) &=& 0\,,
\label{psi2}
\end{eqnarray}
with
\begin{eqnarray}
&&\hspace{-20mm}\Psi_{(+)} (x) = \pmatrix{\rho^A (x)\cr
\lambda^S (x)\cr},
\Psi_{(-)} (x) = \pmatrix{\rho^S (x)\cr
\lambda^A (x)\cr}, \mbox{and}\,\Gamma^\mu =\pmatrix{0 & \gamma^\mu\cr
\gamma^\mu & 0\cr}\nonumber\\
&&
\end{eqnarray}
One can also re-write the equations into the two-component form. Thus, one obtains~\cite{FG} 
equations. Similar formulations have been  presented by~\cite{Markov}, 
and~\cite{Ziino}. The group-theoretical basis for such doubling has been given
in the papers by Gelfand, Tsetlin and Sokolik~\cite{Gelfand}, who first presented 
the theory in the 2-dimensional representation of the inversion group in 1956 (later called as ``the Bargmann-Wightman-Wigner-type quantum field theory" in 1993).

The Lagrangian is
\begin{eqnarray}
&&{\cal L}= \frac{i}{2} \left[\bar \lambda^S \gamma^\mu \partial_\mu \lambda^S - (\partial_\mu \bar \lambda^S ) \gamma^\mu \lambda^S +
\bar \rho^A \gamma^\mu \partial_\mu \rho^A - (\partial_\mu \bar \rho^A ) \gamma^\mu \rho^A +\right.\nonumber\\
&&\left.+\bar \lambda^A \gamma^\mu \partial_\mu \lambda^A - (\partial_\mu \bar \lambda^A ) \gamma^\mu \lambda^A +
\bar \rho^S
\gamma^\mu \partial_\mu \rho^S - (\partial_\mu \bar \rho^S ) \gamma^\mu \rho^S -\right.\nonumber\\
&&\left. - m (\bar\lambda^S \rho^A +\bar \lambda^S \rho^A -\bar\lambda^S \rho^A -\bar\lambda^S \rho^A )
\right ]
\end{eqnarray}

The connection with the Dirac spinors has been found~\cite{Dvoeglazov1,Kirchbach}.\footnote{I also acknowledge
personal communications from D. V. Ahluwalia on these matters.}
For instance,
\begin{eqnarray}
\pmatrix{\lambda^S_\uparrow ({\bf p}) \cr \lambda^S_\downarrow ({\bf p}) \cr
\lambda^A_\uparrow ({\bf p}) \cr \lambda^A_\downarrow ({\bf p})\cr} = {1\over
2} \pmatrix{1 & i & -1 & i\cr -i & 1 & -i & -1\cr 1 & -i & -1 & -i\cr i&
1& i& -1\cr} \pmatrix{u_{+1/2} ({\bf p}) \cr u_{-1/2} ({\bf p}) \cr
v_{+1/2} ({\bf p}) \cr v_{-1/2} ({\bf p})\cr}.\label{connect}
\end{eqnarray}
See also Refs.~\cite{Gelfand,Ziino} and the discussion below. Thus, we can see
that the two 4-spinor sets are connected by the unitary transformations, and this represents
itself the rotation of the spin-parity basis.

The sets of $\lambda$ spinors and of $\rho$ spinors are claimed to be
{\it bi-orthonormal} sets each in the mathematical sense~\cite{Ahluwalia},  provided
that overall phase factors of 2-spinors $\theta_1 +\theta_2 = 0$ or $\pi$.
For instance, on the classical level $\bar \lambda^S_\uparrow
\lambda^S_\downarrow = 2iN^2 \cos ( \theta_1 + \theta_2 )$.\footnote{We used above 
$\theta_1=\theta_2 =0$.} 

Several remarks have been given in the previous works:

\begin{itemize}
\item
While in the massive case there are four $\lambda$-type spinors, two
$\lambda^S$ and two $\lambda^A$ (the $\rho$ spinors are connected by
certain relations with the $\lambda$ spinors for any spin case),  in the
massless case $\lambda^S_\uparrow$ and $\lambda^A_\uparrow$ may identically
vanish, provided that one takes into account that $\phi_L^{\pm 1/2}$ may be
the eigenspinors of ${\bf \sigma}\cdot \hat {\bf n}$, the $2\times 2$ helicity operator.

\item
It was noted that  the possibility exists for generalizations of the concept of the
Fock space, which leads to the ``doubling" Fock space~\cite{Gelfand,Ziino}.

\end{itemize}

It was shown~\cite{Dvoeglazov1} that the covariant derivative (and, hence, the
 interaction) can be introduced in this construct in the following way:
\begin{equation}
\partial_\mu \rightarrow \nabla_\mu = \partial_\mu - ig \L^5 A_\mu\quad,
\end{equation}
where $\L^5 = \mbox{diag} (\gamma^5, \quad -\gamma^5)$, the $8\times 8$
matrix. In other words, with respect to the transformations
\begin{eqnarray}
\lambda^\prime (x)
\rightarrow (\cos \alpha -i\gamma^5 \sin\alpha) \lambda
(x)\quad,\label{g10}\\
\overline \lambda^{\,\prime} (x) \rightarrow
\overline \lambda (x) (\cos \alpha - i\gamma^5
\sin\alpha)\quad,\label{g20}\\
\rho^\prime (x) \rightarrow  (\cos \alpha +
i\gamma^5 \sin\alpha) \rho (x) \quad,\label{g30}\\
\overline \rho^{\,\prime} (x) \rightarrow  \overline \rho (x)
(\cos \alpha + i\gamma^5 \sin\alpha)\quad\label{g40}
\end{eqnarray}
the spinors retain their properties to be self/anti-self charge conjugate
spinors and the proposed Lagrangian~\cite{Dvoeglazov1} remains to be invariant.
This tells us that while self/anti-self charge conjugate states have
zero eigenvalues of the ordinary (scalar) charge operator but they can
possess the axial charge (cf.  with the discussion of~\cite{Ziino} and
the old idea of R. E. Marshak -- they claimed the same).

In fact, from this consideration one can recover the Feynman-Gell-Mann
equation (and its charge-conjugate equation). Our equations can be re-written 
in the two-component form \cite{FG}:
\begin{eqnarray} 
\cases{\left [\pi_\mu^- \pi^{\mu\,-}
-m^2 -{g\over 2} \sigma^{\mu\nu} F_{\mu\nu} \right ] \chi (x)=0\,, &\cr
\left [\pi_\mu^+ \pi^{\mu\,+} -m^2
+{g\over 2} \widetilde\sigma^{\mu\nu} F_{\mu\nu} \right ] \phi (x)
=0\,, &\cr}\label{iii}
\end{eqnarray}
where already one has $\pi_\mu^\pm =
i\partial_\mu \pm gA_\mu$, \, $\sigma^{0i} = -\widetilde\sigma^{0i} =
i\sigma^i$, $\sigma^{ij} = \widetilde\sigma^{ij} = \epsilon_{ijk}
\sigma^k$ and $\nu^{^{DL}} (x) =\mbox{column} (\chi \quad \phi )$.

Next, because the transformations
\begin{eqnarray}
\lambda_S^\prime ({\bf p}) &=& \pmatrix{\Xi &0\cr 0&\Xi} \lambda_S ({\bf p})
\equiv \lambda_A^\ast ({\bf p}),\\
\lambda_S^{\prime\prime} ({\bf p}) &=& \pmatrix{i\Xi &0\cr 0&-i\Xi} \lambda_S
({\bf p}) \equiv -i\lambda_S^\ast ({\bf p}),\\
\lambda_S^{\prime\prime\prime} ({\bf p}) &=& \pmatrix{0& i\Xi\cr
i\Xi &0\cr} \lambda_S ({\bf p}) \equiv i\gamma^0 \lambda_A^\ast
({\bf p}),\\
\lambda_S^{IV} ({\bf p}) &=& \pmatrix{0& \Xi\cr
-\Xi&0\cr} \lambda_S ({\bf p}) \equiv \gamma^0\lambda_S^\ast
({\bf p})
\end{eqnarray}
with the $2\times 2$ matrix $\Xi$ defined as ($\phi$ is the azimuthal
angle  related with ${\bf p}$)
\begin{equation}
\Xi = \pmatrix{e^{i\phi} & 0\cr 0 &
e^{-i\phi}\cr}\quad,\quad \Xi \Lambda_{R,L} ({\bf p} \leftarrow
{\bf 0}) \Xi^{-1} = \Lambda_{R,L}^\ast ({\bf p} \leftarrow
 {\bf 0})\,\,\, ,
\end{equation}
and corresponding transformations for
$\lambda^A$, do {\it not} change the properties of bispinors to be in the
self/anti-self charge-conjugate spaces, the Majorana-like field operator
($b^\dagger \equiv a^\dagger$) admits additional phase (and, in general,
normalization) transformations:
\begin{equation} \nu^{ML\,\,\prime}
(x^\mu) = \left [ c_0 + i({\bf \tau}\cdot  {\bf c}) \right
]\nu^{ML\,\,\dagger} (x^\mu) \,, 
\end{equation} 
where $c_\alpha$ are
arbitrary parameters. The ${\bf \tau}$ matrices are defined over the
field of $2\times 2$ matrices and the Hermitian
conjugation operation is assumed to act on the $c$- numbers as the complex
conjugation. One can parametrize $c_0 = \cos\phi$ and ${\bf c} = {\bf n}
\sin\phi$ and, thus, define the $SU(2)$ group of phase transformations.
One can select the Lagrangian which is composed from the both field
operators (with $\lambda$ spinors and $\rho$ spinors)
and which remains to be
invariant with respect to this kind of transformations.  The conclusion
is: it is permitted the non-Abelian construct which is based on
the spinors of the Lorentz group only (cf. with the old ideas of T. W.
Kibble and R. Utiyama) .  This is not surprising because both the $SU(2)$
group and $U(1)$ group are  the sub-groups of the extended Poincar\'e group
(cf.~\cite{Ryder}).

The Dirac-like and Majorana-like field operators can
be built from both $\lambda^{S,A} ({\bf p})$ and $\rho^{S,A} ({\bf p})$,
or their combinations. For 
instance,
\begin{eqnarray}
&&\nu (x^\mu) \equiv \int {d^3 {\bf p}\over (2\pi)^3} {1\over 2E_p}
\sum_\eta \left [ \lambda^S_\eta ({\bf p}) \, a_\eta ({\bf p}) \,\exp
(-ip\cdot x) +\right.\nonumber\\
&+&\left.\lambda^A_\eta ({\bf p})\, b^\dagger_\eta ({\bf p}) \,\exp
(+ip\cdot x)\right ].\label{oper}
\end{eqnarray}

The anticommutation relations are the following ones (due to the {\it bi-orthonormality}):
\begin{eqnarray}
[a_{\eta{\prime}} ({\bf p}^{\prime}), a_\eta^\dagger ({\bf p}) ]_+ = (2\pi)^3 2E_p \delta ({\bf p} -{\bf p}^\prime) \delta_{\eta,-\eta^\prime}
\end{eqnarray}
and 
\begin{eqnarray}
[b_{\eta{\prime}} ({\bf p}^{\prime}), b_\eta^\dagger ({\bf p}) ]_+ = (2\pi)^3 2E_p \delta ({\bf p} -{\bf p}^\prime) \delta_{\eta,-\eta^\prime}
\end{eqnarray}
Other anticommutators are equal to zero: ($[ a_{\eta^\prime} ({\bf p}^{\prime}), 
b_\eta^\dagger ({\bf p}) ]_+=0$).

Finally, it is interesting to note that
\begin{eqnarray}
\lefteqn{\hspace{-25mm}\left [ \nu^{^{ML}} (x^\mu) + {\cal C} \nu^{^{ML\,\dagger}} (x^\mu) \right
]/2 = \int {d^3 {\bf p} \over (2\pi)^3 } {1\over 2E_p} \sum_\eta \left
[\pmatrix{i\Theta \phi_{_L}^{\ast \, \eta} (p^\mu) \cr 0\cr} a_\eta
(p^\mu)  e^{-ip\cdot x} +\right.}\nonumber \\
&&+\left.\pmatrix{0\cr
\phi_L^\eta (p^\mu)\cr } a_\eta^\dagger (p^\mu) e^{ip\cdot x} \right ]\,
,\\
\lefteqn{\hspace{-25mm}\left [\nu^{^{ML}} (x^\mu) - {\cal C} \nu^{^{ML\,\dagger}} (x^\mu) \right
]/2 = \int {d^3 {\bf p} \over (2\pi)^3 } {1\over 2E_p} \sum_\eta \left
[\pmatrix{0\cr \phi_{_L}^\eta (p^\mu) \cr } a_\eta (p^\mu)  e^{-ip\cdot x}
+\right.}\nonumber\\
&&+\left.\pmatrix{-i\Theta \phi_{_L}^{\ast\, \eta} (p^\mu)\cr 0
\cr } a_\eta^\dagger (p^\mu) e^{ip\cdot x} \right ]\, , 
\end{eqnarray}
thus naturally leading to the Ziino-Barut scheme of massive chiral
fields, Ref.~\cite{Ziino}.

The content of this Section is mainly based on the previous works of the 90s by D. V. Ahluwalia and by me (V. V. Dvoeglazov)
dedicated to the Majorana-like momentum-representation 4-spinors. However, recently the interest to this model raised
again~\cite{Rocha}.

\section{Chirality and Helicity.}

\subsection{History.}

\begin{itemize}

\item
\cite{Ahluwalia} claimed ``Incompatibility 
of Self-Charge Conjugation with Helicity Eignestates and Gauge Interactions". I showed that the gauge interactions 
of $\lambda -$ and $\rho -$ 4-spinors are different. As for the self/anti-self charge-conjugate states
and their relations to helicity eigenstates the question is much more difficult, see below. Either we should accept that the
rotations would have physical significance, or, due to some reasons, we should not apply the equivalence
transformation to the discrete symmetry operators.
As far as I understood~\cite{Ahluwalia} paper,\footnote{He claimed~\cite{Ahluwalia}: ``Just as the operator of parity in the $(j, 0) \oplus (0, j)$ representation space is independent
of which wave equation is under study, similarly the operations of charge conjugation and time
reversal do not depend on a specific wave equation. Within the context of the logical framework
of the present paper, without this being true we would not even know how to define self-/anti self
conjugate $(j, 0) \oplus (0, j)$ spinors."} the latter standpoint is precisely his standpoint.

\item 
Z.-Q. Shi and G. J. Ni  promote a very extreme standpoint. Namely, ``the spin states, the helicity states and the chirality states of fermions in Relativistic Quantum Mechanics are entirely different: a spin state is helicity degenerate; a helicity state can be expanded as a linear combination of the chirality states; the polarization of fermions in flight must be described by the helicity states" (see also the Conclusion Section of the second paper~\cite{Shi}). In fact, they showed experimental consequences of their statement: ``the lifetime
of RH polarized fermions is always greater than of LH ones with the same speed in flight". However, we showed that the helicity, chiral helicity and chirality
operators are connected by the unitary transformations. Do rotations have physical significance in their opinion?

\item
M. Markov wrote long ago {\it two} Dirac equations with  the opposite signs at the mass term~\cite{Markov}. 
\begin{eqnarray}
\left [ i\gamma^\mu \partial_\mu - m \right ]\Psi_1 (x) &=& 0\,,\\
\left [ i\gamma^\mu \partial_\mu + m \right ]\Psi_2 (x) &=& 0\,.
\end{eqnarray}
In fact, he studied all properties of this relativistic quantum model (while he did not know yet the quantum
field theory in 1937). Next, he added and  subtracted these equations. What did he obtain?
\begin{eqnarray}
i\gamma^\mu \partial_\mu \chi (x) - m \eta (x) &=& 0\,,\\
i\gamma^\mu \partial_\mu \eta (x) - m \chi (x) &=& 0\,,
\end{eqnarray}
thus, $\chi$ and $\eta$ solutions can be presented as some superpositions of the Dirac 4-spinors $u-$ and $v-$.
These equations, of course, can be identified with the equations for $\lambda -$ and $\rho -$ we presented above.
As he wrote himself, he was expecting ``new physics" from these equations. 

\item
\cite{SenGupta}  and others claimed that the solutions of the equation (which follows from the general Sakurai
method of derivation of relativistic quantum equations) 
\begin{equation}
\left [
i\gamma^\mu \partial_\mu - m_1 -m_2 \gamma^5 \right ]\Psi = 0\,\label{gd1}
\end{equation}
are {\it not} the eigenstates of chiral [helicity] operator \linebreak $\gamma_0 ({\bf \gamma} \cdot {\bf p})/p$
in the massless limit. The equation may describe both massive and massless $m_1 =\pm m_2$ 
states.
However, in the massive case the equation (\ref{gd1}) has been obtained by the equivalence transformation
of $\gamma$ matrices.

\item
Barut and Ziino~\cite{Ziino} proposed yet another model. They considered
$\gamma^5$ operator as the operator of the charge conjugation. Thus, the charge-conjugated
Dirac equation has the different sign comparing with the ordinary formulation:
\begin{equation}
[i\gamma^\mu \partial_\mu + m] \Psi_{BZ}^c =0\,,
\end{equation}
and the so-defined charge conjugation applies to the whole system, fermion+electromagnetic field, $e\rightarrow -e$
in the covariant derivative. The concept of the doubling of the Fock space has been
developed in Ziino works (cf.~\cite{Gelfand,Dvoeglazov5}). In their case the charge conjugate states
are simultaneously the eigenstates of the chirality.

\end{itemize}

Let us analize the above statements.

\begin{itemize}
 
\item
The helicity operator is:
\begin{equation}
h=\pmatrix{({\bf \sigma}\cdot \hat{\bf p})&0\cr
0&({\bf \sigma}\cdot {\bf p})\cr}
\end{equation}
However, we can do the equivalence transformation of the  helicity $h$-operator by
the unitary matrix. It is known~\cite{Berg}
that one can
\begin{equation}
{\cal U}_1 ({\bf \sigma}\cdot {\bf a}) {\cal U}_1^{-1} = \sigma_3 \vert {\bf a} \vert\,.\label{s3}
\end{equation}
In the case of the momentum vector, one has
\begin{equation}
{\cal U}_1 =\pmatrix{1& p_l/(p+p_3)\cr
-p_r/(p+p_3)&1\cr}
\end{equation}
and
\begin{equation}
U_1 =\pmatrix{{\cal U}_1 &0\cr
0& {\cal U}_1\cr}\,.
\end{equation}
Thus, we obtain:
\begin{equation}
U_1 h  U_1^{-1} = 
\vert {\bf n} \vert \pmatrix{\sigma_3 &0\cr
0&\sigma_3}\,
\end{equation} 
Then, applying other unitary matrix $U_3$:
\begin{eqnarray}
&&\hspace{-20mm}\pmatrix{1&0&0&0\cr
0&0&1&0\cr
0&1&0&0\cr
0&0&0&1\cr} \pmatrix{\sigma_3 &0\cr
0&\sigma_3} \pmatrix{1&0&0&0\cr
0&0&1&0\cr
0&1&0&0\cr
0&0&0&1\cr} = \pmatrix{1&0&0&0\cr
0&1&0&0\cr
0&0&-1&0\cr
0&0&0&-1\cr}=\nonumber\\
&&=\gamma^5_{chiral}\,.
\end{eqnarray}
we transform to the basis, where the helicity is equal to $\gamma^5$, the chirality operator.

\item
\cite{SenGupta} and others introduced the {\it chiral} helicity $\eta =-\gamma_5 h$, which is equal 
(within the sign) to the well-known matrix ${\bf \alpha}$ multiplied by ${\bf n}$. Again,
\begin{eqnarray}
U_1 ({\bf \alpha}\cdot {\bf n}) U_1^{-1} = \vert {\bf n} \vert
\pmatrix{1&0&0&0\cr
0&-1&0&0\cr
0&0&-1&0\cr
0&0&0&1\cr} = \alpha_3 \vert {\bf n}\vert\,.
\end{eqnarray}
with the same matrix
$U_1$.
And applying the second unitary transformation:
\begin{eqnarray}
&&\hspace{-15mm}U_2 \alpha_3 U_2^\dagger =
\pmatrix{1&0&0&0\cr
0&0&0&1\cr
0&0&1&0\cr
0&1&0&0\cr} \alpha_3 \pmatrix{1&0&0&0\cr
0&0&0&1\cr
0&0&1&0\cr
0&1&0&0\cr} = \pmatrix{1&0&0&0\cr
0&1&0&0\cr
0&0&-1&0\cr
0&0&0&-1\cr}=\nonumber\\
&&=\gamma^5_{chiral}\,,
\end{eqnarray}
we again come to the $\gamma_5$ matrix. The determinats are: $Det U_1\neq 0$, $Det U_{2,3}=-1\neq 0$. Thus, helicity, chirality and chiral  helicity are connected by the unitary transformations.

\item
It is {\it not} surprising to have such a situation because different helicity 2-spinors can be connected {\it not only}
by the anti-linear transformation~\cite{Ryder,Ahluwalia} $\xi_h = (-1)^{1/2-h} e^{i\alpha_h} \Theta_{[1/2]} {\cal K} \xi_{-h}$, but the unitary transformation too. For isntance, when we parametrize the 2-spinors as in~\cite{Dv-ff}:
\begin{eqnarray}
\xi_{\uparrow} \,&=&\,N\,e^{i\alpha}\,
\pmatrix{\cos\,(\theta/2)\cr
\sin\,(\theta/2)\,e^{i\,\phi}}\,,\label{exphia}\\
\xi_{\downarrow}\,&=&\,N\,e^{i\beta}\,
\pmatrix{\sin\,(\theta/2)\cr
-\,\cos\,(\theta/2)\,e^{i\,\phi}}\,,\label{exphib}
\end{eqnarray}
we obtain
\begin{equation}
\xi_\downarrow = U\xi_\uparrow = e^{i (\beta -\alpha)} \pmatrix{0& e^{-i\phi}\cr
-e^{i\phi}&0\cr}\xi_\uparrow\,,
\end{equation}
and
\begin{equation}
\xi_\uparrow = U^\dagger \xi_\downarrow = e^{i(\alpha -\beta)} \pmatrix{0& -e^{-i\phi}\cr
e^{i\phi}&0\cr}\xi_\downarrow\,.
\end{equation}

\end{itemize}

To  say that the 4-spinor is the eigenspinor of the {\it chiral helicity}, and, at the same time, it is {\it not!}
the eigenspinor of the helicity operator (and that the physical results would depend on this), 
signifies the same as to say that rotations have physical significance
on the fundamental level.

\subsection{Non-commutativity.}

I present my talk in the ``non-commutativity" section of the QTS7. So, I should also say
a couple of words on the unitarity transformations in that context.

Recently, we analized the Sakurai-van der Waerden method of the derivation of the Dirac equation
(and, the derivation of the higher-spin equations as well)~\cite{Dvoh}. We can start from
\begin{equation}
(E I^{(2)}-{\bf \sigma}\cdot {\bf p}) (E I^{(2)}+ {\bf\sigma}\cdot
{\bf p} ) \Psi_{(2)} = m^2 \Psi_{(2)} \,,
\end{equation}
and obtain
\begin{equation}
[i\gamma_\mu \partial^\mu - m_1 - m_2 \gamma_5] \Psi (x) =0',.
\end{equation}
Alternatively,
\begin{equation}
(E I^{(4)}+{\bf \alpha}\cdot {\bf p} +m\beta) (E I^{(4)}-{\bf\alpha}\cdot
{\bf p} -m\beta ) \Psi_{(4)} =0.\label{f4}
\end{equation}
Of course, as in the original Dirac work, we have
\begin{equation}
\beta^2 = 1\,,\quad
\alpha^i \beta +\beta \alpha^i =0\,,\quad
\alpha^i \alpha^j +\alpha^j \alpha^i =2\delta^{ij} \,.
\end{equation}
For instance, their explicite forms can be chosen 
\begin{equation}
\alpha^i =\pmatrix{\sigma^i & 0\cr
0 & -\sigma^i \cr}\,,\quad
\beta = \pmatrix{0 & 1_{2\times 2}\cr
1_{2\times 2} & 0\cr}\,,
\end{equation}
where $\sigma^i$ are the ordinary Pauli $2\times 2$ matrices.

We can also postulate the non-commutativity for  the sake of general consideraton (as in Ref.~\cite{DV-noncom})
\begin{equation}
[E, {\bf p}^i]_- = \Theta^{0i} = \theta^i\,.
\end{equation}
Therefore the equation (\ref{f4}) will {\it not} lead
to the well-known equation $E^2 -{\bf p}^2 = m^2$. Instead, we have
\begin{equation}
\hspace{-15mm}\left \{ E^2 - E ({\bf \alpha} \cdot {\bf p})
+({\bf \alpha} \cdot {\bf p}) E - {\bf p}^2 - m^2 - i {\bf\sigma}\times I_{(2)}
[{\bf p}\times {\bf p}] \right \} \Psi_{(4)} = 0
\end{equation}
For the sake of simplicity, we may assume the last term to be zero. Thus we come to
\begin{equation}
\left \{ E^2 - {\bf p}^2 - m^2 -  ({\bf \alpha}\cdot {\bf \theta})
\right \} \Psi_{(4)} = 0\,.
\end{equation} 
However, let us make the unitary transformation with $U_1$ matrix (${\bf p} \rightarrow {\bf \theta}$).\footnote{Of course, the certain relations for the components ${\bf \theta}$ should be assumed. Moreover, in our case ${\bf \theta}$ should not depend on $E$ and ${\bf p}$. Otherwise, we must take the noncommutativity $[E, {\bf p}^i]_-$ again.}
For ${\bf \alpha}$ matrices we come to
\begin{eqnarray}
U_1 ({\bf \alpha}\cdot {\bf \theta}) U_1^{-1} = \vert {\bf \theta} \vert
\pmatrix{1&0&0&0\cr
0&-1&0&0\cr
0&0&-1&0\cr
0&0&0&1\cr} = \alpha_3 \vert {\bf\theta}\vert\,.
\end{eqnarray}
And applying the second unitary transformation with $U_2$ matrix as before:
\begin{eqnarray}
\hspace{-15mm}U_2 \alpha_3 U_2^\dagger =
\pmatrix{1&0&0&0\cr
0&0&0&1\cr
0&0&1&0\cr
0&1&0&0\cr} \alpha_3 \pmatrix{1&0&0&0\cr
0&0&0&1\cr
0&0&1&0\cr
0&1&0&0\cr} = \pmatrix{1&0&0&0\cr
0&1&0&0\cr
0&0&-1&0\cr
0&0&0&-1\cr}\,.
\end{eqnarray}
The final equation is
\begin{equation}
[E^2 -{\bf p}^2 -m^2 -\gamma^5_{chiral} \vert {\bf \theta}\vert ] \Psi^\prime_{(4)} = 0\,.
\end{equation}
In the physical sense this implies the mass splitting for a Dirac particle over the non-commutative space. This procedure may be attractive for explanation of the mass creation and the mass splitting for fermions.

\section{Conclusions.}

I. The $(S,0)\oplus (0,S)$ representation space (even in the case of $S=1/2$)
has reacher mathematical structure with manifestation of deep physical consequences, 
which have not yet been  explored before.

\noindent
II. However, several claims made by other researchers concerning with {\it chirality}, {\it helicity}, {\it chiral helicity}
cannot be considered to be true unless someone would confirm that the rotations (unitary transformations)
have physical consequences on the level of the Lorentz-covariant theories.
}

\end{document}